\documentclass{ifacconf}

\usepackage{graphicx}
\usepackage{natbib}
\usepackage{amsmath}
\usepackage{amssymb}
\usepackage[dvipsnames]{xcolor}
\usepackage{tikz}
\usepackage{siunitx}
\usepackage{url}

\newcommand{\obar}[1]{\mkern 1.5mu\overline{\mkern-1.5mu#1\mkern-1.5mu}\mkern 1.5mu}
\newcommand{\ubar}[1]{\mkern 1.5mu\underline{\mkern-1.5mu#1\mkern-1.5mu}\mkern 1.5mu}

\DeclareMathOperator{\Ima}{Im}

\newcommand{\hlgt}[1]{{\color{Green} #1}}

\begin{document}
\begin{frontmatter}

\title{On identifying the non-linear dynamics of a hovercraft using an end-to-end deep learning approach\thanksref{footnoteinfo}} 

\thanks[footnoteinfo]{This research was supported by the Swiss National Science Foundation under the NCCR Automation (grant agreement
51NF40\_180545). Corresponding author: \texttt{roland.schwan@epfl.ch}}

\author[LA,RAO]{R. Schwan} 
\author[LA]{N. Schmid}
\author[LA]{E. Chassaing}
\author[LA]{K. Samaha}
\author[LA]{C. N. Jones}

\address[LA]{Automatic Control Lab, EPFL, Lausanne, Switzerland}
\address[RAO]{Risk Analytics and Optimization Chair, EPFL, Lausanne, Switzerland}

\begin{abstract}
We present the identification of the non-linear dynamics of a novel hovercraft design, employing end-to-end deep learning techniques. Our experimental setup consists of a hovercraft propelled by racing drone propellers mounted on a lightweight foam base, allowing it to float and be controlled freely on an air hockey table. We learn parametrized physics-inspired non-linear models directly from data trajectories, leveraging gradient-based optimization techniques prevalent in machine learning research. The chosen model structure allows us to control the position of the hovercraft precisely on the air hockey table. We then analyze the prediction performance and demonstrate the closed-loop control performance on the real system.
\end{abstract}


\end{frontmatter}

\section{Introduction}

The game of air hockey has been fascinating to robotic researchers all the way back to the '90s, pioneered by the seminal work of \citep{bishop1999} using a link redundant manipulator and vision-based system to play the game of air hockey. This robotic task is especially interesting due to its fast pace and constrained environment.

In this work, we consider a more unconventional method for actuation. While previous work solely focused on hitting the puck using a robotic arm (see recent work by \cite{namiki2013}, \cite{alattar2019}, and \cite{liu2021}), we are using a free-floating, propeller-actuated hovercraft, i.e., an autonomous puck to hit the puck.

This paper introduces the novel hardware architecture and derives a first-principles model suitable for control. In particular, we use an end-to-end deep learning approach, prevalent in the machine learning community, to learn the first-principle model parameters directly from measured trajectories. For this, we adopt a batched learning approach similar to \cite{verhoek2022} allowing us to use established machine learning frameworks.

The contributions of this paper are as follows:
\begin{itemize}
\item We present a novel platform for playing the game of air hockey that serves as a general demonstrator for more advanced control algorithms in the future.
\item We demonstrate a case study on the use of gradient-based modern machine learning tools for the identification of a non-linear system, resulting in a model that enables precise positional control of the hovercraft on an air hockey table.
\item We analyze the effect of structural assumptions and prior modeling knowledge on the resulting prediction and closed-loop control performance.
\item We introduce a control scheme adept at handling delayed motor dynamics and non-unique mappings inherent to the model, demonstrating its efficacy via accurate state reference tracking on the physical system.
\end{itemize}

Datasets and code to reproduce the results is open-source and available online\footnote{\url{https://github.com/PREDICT-EPFL/holohover-sysid}}.

\section{Hardware Design}

\begin{figure}
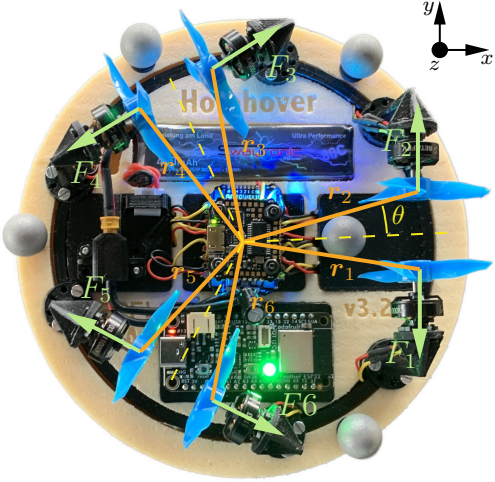

\begin{center}
\include{figures/hovercraft_diagram}
\vspace{-1cm}
\caption{Hardware overview of the hovercraft.} 
\label{fig:hovercraft}
\end{center}
\end{figure}

The hovercraft is $14 \si{cm}$ in diameter and all electronic components are embedded in a $15 \si{mm}$ thick isolation foam base. The propellers are mounted on a 3D-printed ring, which is press-fit into the foam, ensuring that force transmitted to the foam base is evenly distributed to minimize its deformation. The foam base together with all electrical components and the battery only weighs $137 \si{g}$, light enough to float on the air hockey table. A top-down view of the hovercraft can be seen in Figure~\ref{fig:hovercraft}.

The six propellers are powered by brushless motors which are controlled and supplied with power by a \textit{Flywoo GOKU HEX GN405 Nano} running \cite{betaflight2023} which is mounted in the center providing IMU measurements. Under the flight controller, an optical mouse sensor is mounted and provides velocity estimates, which are used for state estimation. The flight controller and mouse sensor are connected to an \textit{ESP32} microcontroller which is running micro-ROS \citep{belsare2023} and is connected to an external computer via Wi-Fi. The software framework including communication is built on the robot operating system (ROS2) \citep{macenski2022}.

\begin{figure}
\begin{center}
\includegraphics[width=8.4cm]{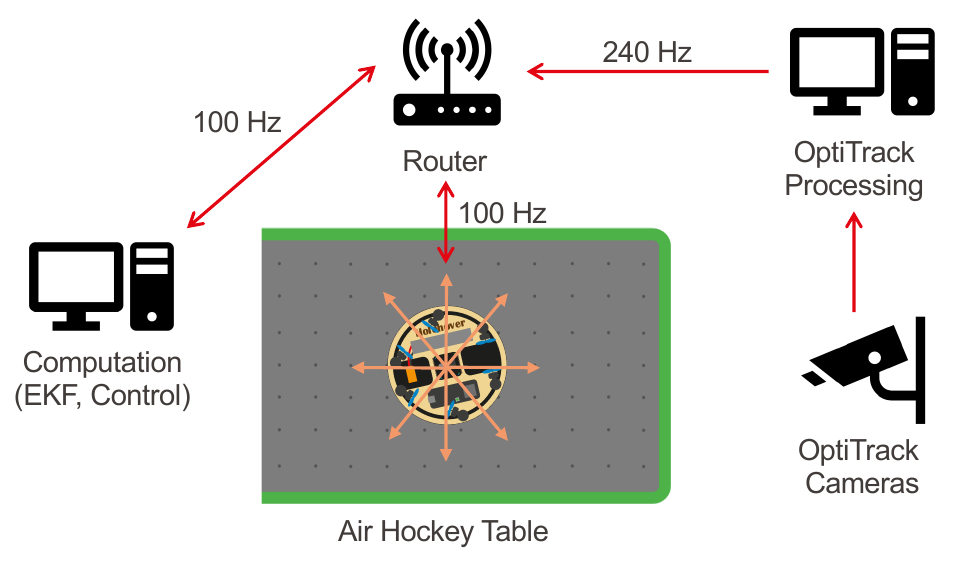}    %
\caption{Schematic overview of the entire system.} 
\label{fig:overview}
\end{center}
\end{figure}

For external position measurements, we utilize an OptiTrack system that consists of multiple infrared cameras and a central processing system. It tracks the infrared markers on the hovercraft at millimeter accuracy at $240 \si{Hz}$. An overview of the entire system is illustrated in Figure~\ref{fig:overview}.

\section{Model} \label{seq:model}

In the following, we derive a first-principles model for the hovercraft. All parameters that are hard to measure, i.e., we want to identify/learn, are \hlgt{highlighted} to indicate that they will become parameters during the identification/learning process. For example, the total mass of the hovercraft can be easily measured, and thus, doesn't need to be identified.

We denote the state of the system as
\begin{equation}
    \boldsymbol{x}=\begin{bmatrix}
        p_x & p_y & \gamma & v_x & v_y & \omega_z
    \end{bmatrix}^T
\end{equation}
where $p_x$ and $p_y$ represent the position, $\gamma \in [-\pi,\pi)$ the orientation, $v_x$ and $v_y$ the velocity, and $\omega_z$ the angular velocity, all in respect to the geometric center.

The control vector, denoted by $\boldsymbol{u} \in [0,1]^6$, represents the signal sent to the motor controllers, with $0$ and $1$ corresponding to no or maximum thrust, respectively.

The evolution of the state is governed by generic rigid body dynamics. In particular, by the following equations:
\begin{equation}
    \boldsymbol{\dot{x}} = \begin{bmatrix}
        v_x & v_y & \omega_z & a_x & a_y & \dot{\omega}_z
    \end{bmatrix}^T
\end{equation}
with
\begin{equation}
    a_x = \frac{F^w_x}{m}, \quad a_y = \frac{F^w_y}{m}, \quad \dot{\omega}_z = \frac{M_z}{\hlgt{I_z}},
\end{equation}
where $F^w_x$ and $F^w_y$ are the $x$ and $y$ components of the forces in world-frame, $m$ is the mass of the system, and $M_z$ the moment acting on the geometric center with inertia \hlgt{$I_z$}. Estimating the inertia directly can be a tough task due to all the electronic components.

The position of each motor in body-frame is given by
\begin{equation}
\begin{aligned}
    \boldsymbol{r}_{2i+1} &= \begin{bmatrix}
        d \cos(\frac{\pi i}{3} - \delta + \hlgt{\alpha}) &
        d \sin(\frac{\pi i}{3} - \delta + \hlgt{\alpha}) &
        0
    \end{bmatrix}^T, \\
    \boldsymbol{r}_{2i+2} &= \begin{bmatrix}
        d \cos(\frac{\pi i}{3} + \delta + \hlgt{\alpha}) &
        d \sin(\frac{\pi i}{3} + \delta + \hlgt{\alpha}) &
        0
    \end{bmatrix}^T,
\end{aligned}
\end{equation}
for $i=0,1,2$, where $d=4.65$~cm is the distance from the center of each motor, $\hlgt{\alpha}$ a rotational offset (nominally~$0^\circ$) due to misalignment of the tracking system, and $\delta = 18.8^{\circ}$ the angle between the center of a propeller pair and the propellers itself. Note that we model everything on the $x$-$y$-plane. Otherwise, we would get moments outside the $z$-axis contradicting the planar-constrained nature of the hovercraft. 

The force vectors of each propeller in body-frame are given by
\begin{equation}
    \boldsymbol{F}^b_{2i+1} = F_{2i} \boldsymbol{v}_{i+1}, \quad
    \boldsymbol{F}^b_{2i+2} = -F_{2i+1} \boldsymbol{v}_{i+1},
\end{equation}
with $\boldsymbol{v}_{i+1} = \begin{bmatrix} -\sin(\frac{\pi i}{3}) & \cos(\frac{\pi i}{3}) & 0 \end{bmatrix}^T$ for $i=0,1,2$,\\
where $F_i$ is the non-linear mapping from signal $u_i$ to the magnitude of thrust of the propeller $i$, which is further discussed in Section~\ref{subsec:thrust_model}.

This gives us the moment
\begin{equation}
    \boldsymbol{M}=\sum_{i=1}^6 \boldsymbol{r}_i \times \boldsymbol{F}^b_i,
\end{equation}
and the force in world-frame
\begin{equation}
    \boldsymbol{F}^w= R^{wb}(\gamma) \sum_{i=1}^6 \boldsymbol{F}^b_i,
\end{equation}
where the rotation matrix
\begin{equation}
    R^{wb}(\gamma) = \begin{bmatrix}
        \cos(\gamma) & -\sin(\gamma) & 0 \\
        \sin(\gamma) & \cos(\gamma) & 0 \\
        0 & 0 & 1
    \end{bmatrix},
\end{equation}
transforms a vector from body to world-frame.

\subsection{Corrections due to shifted center of mass}

The model described in the previous section is only valid under the assumption that the center of mass aligns with the geometric center of the disc, which we want to control. Thus, given an unaligned center of mass \hlgt{$\boldsymbol{r}^{CoM}$}, the moment around the center of mass becomes
\begin{equation}
    \boldsymbol{M}^{CoM}=\sum_{i=1}^6 (\boldsymbol{r}_i - \hlgt{\boldsymbol{r}^{CoM}}) \times \boldsymbol{F}^b_i,
\end{equation}
and since the angular acceleration is the same across the entire ridged body, we get
\begin{equation}
    \dot{\omega}_z = \frac{M^{CoM}_z}{\hlgt{I^{CoM}_z}},
\end{equation}
where \hlgt{$I^{CoM}_z$} is now the inertia around the center of mass. With this, we can correct the linear acceleration as
\begin{equation} \label{eq:acc_corr}
    \boldsymbol{a} = \frac{\boldsymbol{F}^w}{m} + \hlgt{\boldsymbol{r}^{CoM}} \times \boldsymbol{\dot{\omega}} + \boldsymbol{\omega} \times \left( \hlgt{\boldsymbol{r}^{CoM}} \times \boldsymbol{\omega} \right).
\end{equation}

\subsection{Thrust model} \label{subsec:thrust_model}

In \cite{pounds2002}, the thrust of each motor propeller is modeled as
\begin{equation}
    F_i = \kappa \dot{\varphi}_i^2, \quad \forall i=1,\dots,6,
\end{equation}
where $\kappa$ is a proportionality constant and $\dot{\varphi}_i$ are the individual motor velocities. To account for potential aerodynamic effects, we relax this model and parameterize it with a 3rd-order polynomial with zero constant for each motor, due to no force generation if the motor is not spinning. We tested different polynomials up to degree five but found a 3rd-order polynomial works best for our system.
\begin{equation} \label{eq:thrust_poly}
    F_i = \hlgt{a_i} \dot{\varphi}_i + \hlgt{b_i} \dot{\varphi}_i^2 + \hlgt{c_i} \dot{\varphi}_i^3, \quad \forall i=1,\dots,6.
\end{equation}
The motor velocity is controlled through \cite{betaflight2023} which takes our input signal $\boldsymbol{u}$ and transmits it to the electronic speed controller which powers the motors. Similar to \cite{pounds2002}, we can model the propeller velocities as a first-order model
\begin{equation} \label{eq:motor_dynamics}
    \boldsymbol{\ddot{\varphi}} = \frac{\boldsymbol{u} - \boldsymbol{\dot{\varphi}}}{\hlgt{\tau}},
\end{equation}
with time constant $\hlgt{\tau} > 0$. Note that we have no measurements of the motor velocities, i.e., the scaling of $\boldsymbol{\dot{\varphi}}$ does not correspond to the real motor velocities. Thus, instead of scaling the input $\boldsymbol{u}$ such that $\boldsymbol{\dot{\varphi}}$ would match the real motor velocities, we can also scale the polynomial coefficients in \eqref{eq:thrust_poly} instead to avoid an over parameterization of the model.

Hence, to accommodate for the motor dynamics, we extend the state space as
\begin{equation} \label{eq:extended_model_dynamics}
    \boldsymbol{z}=\begin{bmatrix}
        \boldsymbol{x} \\ \boldsymbol{\dot{\varphi}}
    \end{bmatrix}, \quad \boldsymbol{\dot{z}} = f(\boldsymbol{z}, \boldsymbol{u}),
\end{equation}
resulting in our complete first-principles model.

\subsection{Representation through a configuration matrix} \label{subsec:configuration_matrix}

We can compactly describe the non-linear portion of the dynamics as
\begin{equation} \label{eq:acc_dyn}
    \begin{bmatrix}
        a_x \\ a_y \\ \dot{\omega}_z
    \end{bmatrix} = \begin{bmatrix}
        1 & 0 & \hlgt{r^{CoM}_y} \\
        0 & 1 & \hlgt{-r^{CoM}_x} \\
        0 & 0 & 1
    \end{bmatrix} R^{wb}(\gamma)\Pi(\hlgt{\theta})F(\boldsymbol{\dot{\varphi}}) + g(\omega_z),
\end{equation}
where the entries in the first matrix are due to the second term in \eqref{eq:acc_corr}, $\Pi(\hlgt{\theta}) \in \mathbb{R}^{3 \times 6}$ is a state-independent dense matrix depending on learnable parameters $\hlgt{\theta}$, $F \colon [0,1]^6 \to \mathrm{R}^6$ is the thrust mapping \eqref{eq:thrust_poly}, and $g(\omega_z) = \hlgt{\boldsymbol{r}^{CoM}} \omega_z^2$ is the last term in \eqref{eq:acc_corr}.

Since $\Pi(\hlgt{\theta})$ is state-independent, it can be seen as a configuration matrix that maps the forces to linear and angular accelerations in body-frame with respect to the center of mass. Hence, instead of parametrizing $\Pi(\hlgt{\theta})$ as described in the previous sections analytically, we can also directly learn the matrix $\hlgt{\Pi}$. We compare the difference between the first-principles model and the configuration matrix model in Section~\ref{sec:results}.

\section{Control} \label{seq:control}

For control, we consider the system as a point mass and take the linear and angular accelerations as virtual inputs~$\boldsymbol{v}$, giving us the discretized dynamics
\begin{align}
    \boldsymbol{x}_{k+1} &= A \boldsymbol{x}_k + B \boldsymbol{v}_k, \label{eq:point_mass_subsystem} \\
    \boldsymbol{v}_k & = M(\gamma_k)F(\boldsymbol{\dot{\varphi}}_k) + g((\omega_z)_k), \\
    \boldsymbol{\dot{\varphi}}_{k+1} &= A^\varphi \boldsymbol{\dot{\varphi}}_k + B^\varphi \boldsymbol{u}_k, \label{eq:dis_motor_dynamics}
\end{align}
with $M(\gamma) \in \mathrm{R}^{3 \times 6}$ combining the matrix terms in \eqref{eq:acc_dyn}, and respective matrices $A \in \mathrm{R}^{6 \times 6}$, $B \in \mathrm{R}^{6 \times 3}$, and diagonal matrices $A^\varphi \in \mathrm{R}^{6 \times 6}$ and $B^\varphi \in \mathrm{R}^{6 \times 6}$ which can be exactly calculated as described in \cite[Chapter 14]{decarlo1989}. Note that this discretization is not entirely correct, since $\boldsymbol{v}_k$ is changing for a constant $\boldsymbol{\dot{\varphi}}_k$ if the hovercraft is rotating. However, the error is small assuming we sample fast enough and $\omega_z$ is kept small.

We now can easily design a controller $\boldsymbol{v}_k = \kappa(\boldsymbol{x}_k)$ for the linear subsystem \eqref{eq:point_mass_subsystem} using, for example, linear quadratic regulator (LQR) synthesis or model predictive control. %
The only challenging remaining part is the mapping from $\boldsymbol{v}_k$ to $\boldsymbol{u}_k$, which includes a non-linearity, non-unique mapping, and the time delay introduced in the motor dynamics \eqref{eq:dis_motor_dynamics}.

Since at time step $k$, the motor velocity $\boldsymbol{\dot{\varphi}}_k$ is already determined, input $\boldsymbol{u}_k$ has no influence at $\boldsymbol{v}_k$ anymore, i.e., $\boldsymbol{v}_k$ is only dependent on $\boldsymbol{\dot{\varphi}}_k$. Hence, we calculate the virtual input $\boldsymbol{v}_{k+1}=\kappa(A \boldsymbol{x}_k + B \boldsymbol{v}_k)$ for the next time step instead.

In Section~\ref{subsec:reg_force_model}, we enforce $F(\cdot)$ to be a monotonically increasing polynomial. To find $\boldsymbol{\dot{\varphi}}_{k+1}$ and consequently $\boldsymbol{u}_k$, we use the fact that we can map from a desired thrust back to a motor velocity, which we denote by the function $F^{-1} \colon \Ima F \to [0,1]^6$. Hence, we can find the inverse easily by applying a couple of Newton steps.

We find the mapping from $\boldsymbol{v}_{k+1}$ to a thrust vector $\boldsymbol{F}_{k+1}$ by solving the quadratic program (QP)
\begin{equation} \label{eq:force_qp}
\begin{aligned}
    \min_{\boldsymbol{F}_{k+1},s_1,s_2} & \|\boldsymbol{F}_{k+1}\|_2^2 + \mu (1-s_1)^2 + \mu (1-s_2)^2 \\
    \text{s.t.} & \begin{bmatrix}
        s_1 & 0 & 0 \\
        0 & s_1 & 0 \\
        0 & 0 & s_2
    \end{bmatrix} \boldsymbol{v}_{k+1} = M(\gamma_{k+1})\boldsymbol{F}_{k+1} + g((\omega_z)_{k+1}), \\
    & \ubar{\boldsymbol{F}}_{k+1} = F(A^\varphi \boldsymbol{\dot{\varphi}}_k + B^\varphi \boldsymbol{u}_\text{min}), \\
    & \obar{\boldsymbol{F}}_{k+1} = F(A^\varphi \boldsymbol{\dot{\varphi}}_k + B^\varphi \boldsymbol{u}_\text{max}), \\
    & \ubar{\boldsymbol{F}}_{k+1} \leq \boldsymbol{F}_{k+1} \leq \obar{\boldsymbol{F}}_{k+1}, \quad\quad 0 \leq s_1,s_2 \leq 1,
\end{aligned}
\end{equation}
where $\mu \gg 1$ is a penalty weight for the slack variables $s_1$ and $s_2$. The control input is then given by
\begin{equation}
    \boldsymbol{u}_k = (B^\varphi)^{-1}(F^{-1}(\boldsymbol{F}_{k+1}) - A^\varphi \boldsymbol{\dot{\varphi}}_k).
\end{equation}
The slack variables $s_1,s_2\in [0,1]$ in \eqref{eq:force_qp} ensure that the problem is always feasible while ensuring $\boldsymbol{u}_k \in [\boldsymbol{u}_\text{min}, \boldsymbol{u}_\text{max}]$. If the thrust constraints can be met, the large penalty parameter $\mu$ ensures that $s_1,s_2=1$, otherwise, if the desired accelerations $v_k$ can not be met, the linear accelerations are scaled down at the same ratio due to the shared slack variable $s_1$.

For our practical implementation used in Section~\ref{sec:experiment}, the virtual controller $\kappa(\boldsymbol{x}_k)$ was chosen to be a hand-tuned LQR controller. We use the QP solver PIQP \citep{schwan2023} to solve problem \eqref{eq:force_qp} reliably in real-time at well over 100~Hz.

\section{Learning from data}

\begin{figure}
\begin{center}
\includegraphics[width=8.4cm]{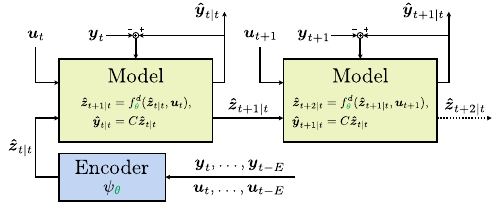}    %
\vspace{-3mm}
\caption{Schematic overview of learning architecture.} 
\label{fig:learning_architecture}
\end{center}
\end{figure}

Consider the data-generating discrete-time system given by
\begin{equation}
\Sigma:\left\{\begin{aligned}
\boldsymbol{z}_{k+1} & = f^d\left(\boldsymbol{z}_k, \boldsymbol{u}_k\right), \\
\boldsymbol{y}_k & = C \boldsymbol{z}_k,
\end{aligned}\right.
\end{equation}
with sampling period $dt$, where $f^d\left(\boldsymbol{z}_k, \boldsymbol{u}_k\right)$ represents the real discrete-time dynamics of the underlying system, and $C\in \mathbb{R}^{3 \times 12}$ is a matrix selecting the measured variables $\boldsymbol{y}=(p_x, p_y, \gamma)$ from the state.

\subsection{Minimizing batched prediction loss}

Assume we have a dataset $\mathcal{D}_N = \{\boldsymbol{y}_k,\boldsymbol{u}_k\}_{k=1}^N$ generated from $\Sigma$. We want to find the parameters $\hlgt{\theta}$ of our model such that the loss function
\begin{equation} \label{eq:loss_full}
    \mathcal{L}_{\mathcal{D}_N}(\hlgt{\theta}) = \frac{1}{N} \sum_{k=1}^N  \|\boldsymbol{\hat{y}}_{k} - \boldsymbol{y}_{k}\|_2^2
\end{equation}
is minimized, where $\boldsymbol{\hat{y}}_{k}$ is the prediction of our model at time step $k$.

Predicting over the whole dataset and minimizing \eqref{eq:loss_full} directly can be rather non-smooth and sensitive to local minima, as outlined in \cite{ribeiro2020}. Thus, we adopt a batched learning procedure similar to \cite{verhoek2022}. In particular, instead of predicting over the whole dataset length, we only predict over subsections, giving us the following batched minimization problem
\begin{equation}
\begin{aligned}
    \min_{\hlgt{\theta}} \mathcal{L}_{\mathcal{D}_N}^{\text{bat}}(\hlgt{\theta}) &= \frac{1}{T \cdot N_\text{batch}} \sum_{t \in \mathcal{I}} \sum_{k=0}^{T-1} \|\boldsymbol{\hat{y}}_{t+k|t} - \boldsymbol{y}_{t+k}\|_2^2 \\
    \boldsymbol{\hat{z}}_{t+k+1|t} &= f^{d}_{\hlgt{\theta}}(\boldsymbol{\hat{z}}_{t+k|t}, \boldsymbol{u}_{t+k}), \\
    \boldsymbol{\hat{y}}_{t+k|t} &= C\boldsymbol{\hat{z}}_{t+k|t}, \\
    \boldsymbol{\hat{z}}_{t|t} &= \psi_{\hlgt{\theta}}(\boldsymbol{y}_{t},\dots,\boldsymbol{y}_{t-E},\boldsymbol{u}_{t},\dots,\boldsymbol{u}_{t-E}),
\end{aligned}
\end{equation}
where $\mathcal{I} \subseteq \{1, \dots, N-T-E+1\}$ is the index set for the training batches with $N_{\text{batch}}=|\mathcal{I}|$, $f^{d}_{\hlgt{\theta}}(\cdot)$ the discretized model with parameters $\hlgt{\theta}$, and $\psi_{\hlgt{\theta}}(\cdot)$ is the encoder function giving an estimate of the extended state given past measurements and inputs. $T$ is the prediction length and $E$ is the encoder length, i.e., the trajectory length the encoder estimates the current state $\boldsymbol{\hat{z}}_{t|t}$ from. The $|$ notation indicates the difference between the local prediction index on the left and the global starting index of the data on the right. Figure~\ref{fig:learning_architecture} shows an overview of the learning architecture.

The discretized model $f^{d}_{\hlgt{\theta}}(\cdot)$ is obtained by applying an explicit Runge-Kutta method of fourth order to the continuous system dynamics \eqref{eq:extended_model_dynamics}. While adaptive integration schemes are possible in a differentiable framework as popularized by \cite{chen2018neuralode}, they incur a higher computational cost.

\subsection{Encoder design} \label{subsec:encoder}

Given that the extended state is not directly observable, we must infer $\boldsymbol{z}_{t|t}$ from past measurements. While deep neural networks offer a potential method for state estimation as suggested by \cite{beintema2023}, their complexity and lack of transparency render them an unsuitable choice for our specific context. Therefore, we opt for a more traditional and straightforward technique that better serves the requirements of our analysis.

The velocity components of $\hat{z}_{t|t}$, i.e., $v_x$, $v_y$, and $\omega_z$ at time $t$, are determined during an initial processing phase as delineated in Section~\ref{subsec:data_preprocessing}. In the absence of direct motor velocity measurements $\boldsymbol{\dot{\varphi}}$, we initiate $\boldsymbol{\dot{\varphi}}_{t-E|t}$ with $\boldsymbol{u}_{t-E}$ and integrate them forward in time using the motor dynamics \eqref{eq:motor_dynamics}. Provided the encoder length is sufficiently long, the estimate $\boldsymbol{\dot{\varphi}}_{t|t}$ will approach the actual velocity, thanks to the inherent stability of the motor dynamics. It is important to note that the encoder function $\psi_{\hlgt{\theta}}(\cdot)$ is influenced by parameters, specifically the time constant $\hlgt{\tau}$.

\subsection{Regularization of force model} \label{subsec:reg_force_model}

We want to incorporate the physical knowledge that thrust \eqref{eq:thrust_poly} has to be monotonically increasing for $\dot{\varphi}_i \in [0,1]$ for propeller $i$, i.e., the derivative of \eqref{eq:thrust_poly} has to be positive for $\dot{\varphi}_i \in [0,1]$, which is equivalent to condition
\begin{equation} \label{eq:thrust_cond}
\begin{aligned}
    0 \leq \min_{\dot{\varphi}_i} &\; \hlgt{a_i} + 2\hlgt{b_i} \dot{\varphi}_i + 3\hlgt{c_i} \dot{\varphi}_i^2 \\
    \text{s.t.} &\; 0 \leq \dot{\varphi}_i \leq 1.
\end{aligned}
\end{equation}
We can prove the following proposition, which gives a set of conditions which have to hold:
\begin{prop} \label{prop:thrust}
The condition \eqref{eq:thrust_cond} holds if $\hlgt{a_i} \geq 0$, $\hlgt{a_i} + 2\hlgt{b_i} + 3\hlgt{c_i} \geq 0$, and $\hlgt{a_i} - \frac{\hlgt{b_i}^2}{3\hlgt{c_i}} \geq 0$ for $\hlgt{c_i} > 0$ and $-\frac{\hlgt{b_i}}{3\hlgt{c_i}}\in [0,1]$.
\end{prop}
\begin{pf}
Condition \eqref{eq:thrust_cond} has to hold at the boundaries of $[0,1]$ giving us the first two conditions. Due to continuity, \eqref{eq:thrust_cond} is true if $\hlgt{c_i} \neq 0$ and the extrema point, which is located at $\dot{\varphi}_i^\text{ext}=-\frac{\hlgt{b_i}}{3\hlgt{c_i}}$, is in $[0,1]$, and objective $\hlgt{a_i} - \frac{\hlgt{b_i}^2}{3\hlgt{c_i}} > 0$. Is $\dot{\varphi}_i^\text{ext} \notin [0,1]$, the sign of the slope of the objective in $[0,1]$ is not switching, implying that \eqref{eq:thrust_cond} already holds due to the boundary conditions. Thus, this gives us the last condition in the proposition.
\end{pf}

Using Proposition~\ref{prop:thrust}, we can design a regularization term which we add to the loss such that the thrust is monotonically increasing. In particular, we add the regularizer
\begin{equation}
\begin{aligned}
    \mathcal{L}^{\text{reg}}(\hlgt{\theta}) = w_\text{reg} \sum_{i=1}^{6} ( & \max(0, -\hlgt{a_i}) \\
    &+ \max(0, -\hlgt{a_i} - 2\hlgt{b_i} - 3\hlgt{c_i}) \\
    &+ \max(0, \frac{\hlgt{b_i}^2}{3\hlgt{c_i}} - \hlgt{a_i}) ),
\end{aligned}
\end{equation}
where $w_\text{reg} > 0$ is the regularization weight, and the last term is conditionally added if $\hlgt{c_i} > 0$ and $-\frac{\hlgt{b_i}}{3\hlgt{c_i}}\in [0,1]$. Thus, we minimize the combined loss
\begin{equation} \label{eq:combined_loss}
    \mathcal{L}(\hlgt{\theta}) = \mathcal{L}_{\mathcal{D}_N}^{\text{bat}}(\hlgt{\theta}) + \mathcal{L}^{\text{reg}}(\hlgt{\theta}).
\end{equation}
We found that the obtained trust profiles exhibit a non-monotonic behavior without the regularization term, emphasizing the regularization's importance to achieve physically accurate models.

\section{Experimental setup} \label{sec:experiment}

For the experiment design, we apply random signals to the system. In particular, the input signal should span the whole input space such that we can identify the thrust model \eqref{eq:thrust_poly} properly. But applying random signals leads to the hovercraft hitting the walls of the air hockey table. Hence, we adopt a hybrid approach of combining the LQR controller from Section~\ref{seq:control} with random signals.

For this, we uniformly sample signal pairs $\boldsymbol{s} \in [-1,1]^3$ which are added to the input signals
\begin{equation}
\begin{aligned}
    u_{2i+1} &=\min(1.0, \max(u_\text{min},u^c_{2i+1} + s_i)), \\
    u_{2i+2} &=\min(1.0, \max(u_\text{min},u^c_{2i+2} + s_i)), \\
\end{aligned}
\end{equation}
for $i=1,2,3$ with $\boldsymbol{u}^c=\kappa(\boldsymbol{x})$ being the controller signal and $u_\text{min}=0.03$. This ensures that only one motor per propeller pair is producing thrust, while the minimum signal $u_\text{min}$ ensures that the motor keeps spinning. It was observed that the startup sequence of the motors can be very inconsistent and introduces a high delay. The random signal $\boldsymbol{s}$ is then kept constant for $200 \si{ms}$, i.e., for $20$ time steps with a controller frequency of $100 \si{Hz}$.

\subsection{Data preprocessing} \label{subsec:data_preprocessing}

Before using the data for identification, we preprocess the experiment data to correct for timestamp misalignment, calculating velocity and acceleration estimates, and transmission delays.

\textbf{Data interpolation}: Control signals that are sent at $100\si{Hz}$ to the hovercraft are not aligned with position measurements coming from the OptiTrack system at $240\si{Hz}$. Thus, we interpolate and resample the control signals at $240\si{Hz}$ with a zero-order hold.

\textbf{Velocity and acceleration estimates}: We calculate velocity and acceleration estimates from the OptiTrack data using a numerical finite difference. To smooth the resulting noisy signals, we use a Savitzky-Golay filter.

\textbf{Delay compensation}: Between the position measurements and the control signals exists a delay due to network delays. To find and correct the delay, we calculate the accelerations using our model \eqref{eq:acc_dyn} and correlate it against the accelerations obtained from the finite difference estimates. We then shift the control signals accordingly. Due to the resulting changes, we correlate the signals again and iterate until no more shifts are necessary. The resulting total shift is 5 time steps, corresponding to $20.8\si{ms}$ in delay.

\subsection{Models}

We are identifying and comparing five different models, more specifically, we are considering models with an increasing number of free parameters which are described in Table~\ref{table:models}. $\hlgt{\tau}$ equal $0$ indicates no modeling of the first-order model \eqref{eq:motor_dynamics}, i.e., the motor speeds are assumed to equal the input signal.

Models M0, M1, and M2 investigate the sensitivity of the inertia and motor time constant parameters, whereas in model M2 the inertia is fixed to the inertia of a homogenous disk. Models M3 and M4 compare the difference between the first-principles model and the representation through the configuration matrix, as discussed in Section~\ref{subsec:configuration_matrix}.

\begin{table}%
\centering
\begin{tabular}{c c c c c c c}
 \hline
 & \\ [-2.5ex]
 Model & \hlgt{\eqref{eq:thrust_poly}} & \hlgt{$I^{CoM}_z$} & \hlgt{$\tau$} & \hlgt{$\alpha$} & \hlgt{$\boldsymbol{r}^{CoM}$} &  \hlgt{$\Pi$}  \\
 \hline
 & \\ [-2.5ex]
 M0 & free & $0.337 \cdot 10^{-3}$ & $0$ & $0$ & $0$ & - \\
 M1 & free & free & $0$ & $0$ & $0$ & - \\
 M2 & free & $0.337 \cdot 10^{-3}$ & free & $0$ & $0$ & - \\
 M3 & free & free & free & free & free & - \\
 M4 & free & - & free & - & - & free \\
 \hline
\end{tabular}
\vspace{2mm}
\caption{Free and fixed parameters for evaluated models. `-' indicates unused parameters.}
\label{table:models}
\end{table}

\subsection{Learning}

For training, we use the machine learning framework PyTorch. We run two independent experiments, recording $90 \si{s}$ worth of data each. The first experiment is exclusively used for training and the second for validation, i.e. all results reported are solely based on the validation data set.

For each model, we minimize the loss $\mathcal{L}(\hlgt{\theta})$ as described in Section~\ref{subsec:reg_force_model}. We use Adam \citep{kingma2015} with a learning rate of $10^{-3}$ and run the learning procedure for $150$ epochs per model with a batch size of $512$. The prediction and encoder lengths were chosen to be $E=20$ and $T=50$, corresponding to an encoder window of $83 \si{ms}$ and a prediction of $0.208 \si{s}$, respectively.

\section{Results} \label{sec:results}

\begin{table}[ht]
\centering
\begin{tabular}{c c c c} 
 \hline
 & \\ [-2.5ex]
 Model & $x$ ($\si{mm}$) & $y$ ($\si{mm}$) & $\gamma$ ($\si{rad}$) \\
 \hline
 & \\ [-2.5ex]
 M0 & $16.2$ & $15.7$ & $0.501$ \\
 M1 & $15.1$ & $15.0$ & $0.495$ \\
 M2 & $12.9$ & $11.4$ & $0.391$ \\
 M3 & $10.8$ & $9.8$ & $0.387$ \\
 M4 & $10.8$ & $9.6$ & $0.387$ \\
 \hline
\end{tabular}
\vspace{2mm}
\caption{RMSE prediction error after $0.2 \si{s}$ on validation dataset.}
\label{table:prediction_error}
\vspace{-3mm}
\end{table}

We evaluate the prediction performance of each model on the validation data set by dividing it into non-overlapping trajectories of length $T+E$, evaluate the encoder $\psi_{\hlgt{\theta}}(\cdot)$ on the first $E$ data points, and use the model to integrate the model for $T$ steps. 

In Table~\ref{table:prediction_error} we show the root mean squared error (RMSE) of the prediction compared against the OptiTrack measurements at a prediction horizon of $48$ time steps or equivalently $0.2 \si{s}$. As expected, the prediction error decreases with more expressive models. Comparing M1 and M2, we can also see that the first-order motor dynamics have a higher impact than learning the correct inertia.

Comparing M3 and M4 we can see that learning the configuration matrix is on par with the full first-principles model. Comparing the thrust model polynomials \eqref{eq:thrust_poly} between M3 and M4 reveals that the max thrust predicted by M4 is considerably larger than that of M3, suggesting that the configuration matrix \hlgt{$\Pi$} is compensating, resulting in a wrongly scaled thrust profile. Considering that we lose interpretability when using the configuration matrix formulation, it might be better to implement the first-principles model. For example, this also allows us to accurately visualize the correct thrust without scaling issues.

\subsection{Reference point tracking}

\begin{figure}
\begin{center}
\includegraphics[width=8.4cm]{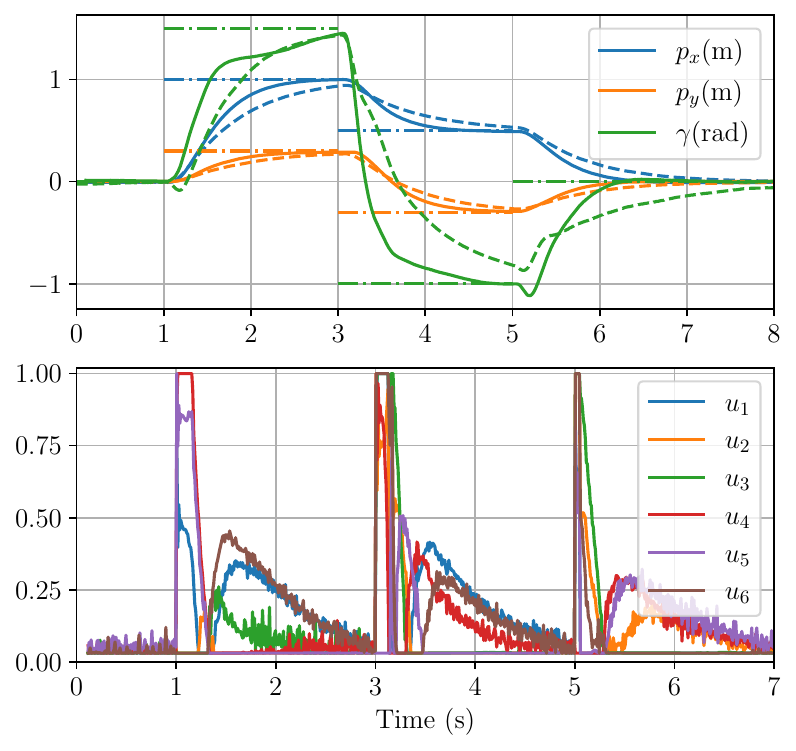}    %
\vspace{-3mm}
\caption{Trajectories of reference tracking controller for position and orientation for models M0 (dashed) and M4 (solid) for three set point changes (top), with its corresponding control inputs for model M4 (bottom). The set points are indicated by the dot-dashed lines.}
\label{fig:control_lqr}
\end{center}
\end{figure}

To demonstrate the applicability of the identified model, we apply it in closed-loop using an LQR controller as discussed in Section~\ref{seq:control} for reference point tracking. We implement models M0 and M4 and apply three set point changes using the same controller tuning. The resulting closed-loop trajectories can be seen in Figure~\ref{fig:control_lqr}. Note the noticeably worse control performance for model M0, resulting in steady-state offsets. Using the more accurate model M4, we achieve better tracking, although due to input saturations enforced by the QP \eqref{eq:force_qp}, we get some non-linear behaviors that can be specifically observed after set point changes in the orientation of the hovercraft.

\section{Conclusion}

We have presented the identification of a non-linear model of a novel hovercraft architecture, employing an end-to-end learning approach. We showed the prediction capabilities of the model and outlined a control strategy compensating for the first-order motor dynamics.

\subsection{Future directions}

In this work, we kept the control architecture simple, utilizing a standard LQR controller, but we are planning to deploy more elaborate control schemes in the future, including model predictive control to adhere to input and state constraints, and time-optimal control. The latter, especially, is necessary to achieve the long-term goal of playing air hockey against humans autonomously.

\bibliography{refs}

\end{document}